# Economic and technical study for the construction of a 1 MW grid-connected solar power plant in southern Iran

Mahmoud Makkiabadi[1]

[1]Department of mechanical engineering, Amirkabir University of Technology, Tehran, Iran

(Mahmud.makiabadi@gmail.com)

**Abstract**

Renewable energy such as solar and wind energy can solve the major problems of humanity such as electricity and fresh water. The renewable energy sources are promising to take a significant share in the energy sector as a viable option for integration with conventional fossil fuel power plants. In this paper, the production of 1 MW of electricity for several households in the city of Sirjan in southern Iran has been studied. The actual data required by the model including solar irradiation, air temperature, load profile, cost of energy for Sirjan, Iran have been utilized in the proposed model. Considering the Iranian market, the fixed and current costs of building this power plant have been studied. Then, using Hummer software, the amount of electricity production per month has been studied. Results showed that Scaled data were used for calculations in HOMER. It had a scaled annual average of 1127 kWh/day and the peak load was 0.467 kW. The maximum electricity energy is obtained in July.

## Introduction

By reducing the supply of fossil fuels such as oil and gas in the coming years, humans will have to build a solar power plant to power themselves [1-2]. Commonly hybrid energy systems use solar, wind, and hydro energy sources, although most of the renewable energy available on earth consists of different forms of solar energy [3-5]. Iran, with its high ability to receive solar energy and also a large area, has a potential advantage in the construction of solar power plants [6]. Hybrid energy, which is the use of different kinds of energy, is more efficient than conventional energy generation [7]. The availability of wind energy in Colombia, combined with biomass energy, has had a significant influence on the Caribbean region [8]. The exploitation of this source of energy can be an excellent solution to the energy problems prevalent in the region for solving the load flows problem such as congestion, and load flow control [9-11]. This solution lies in the design of a hybrid renewable energy plant that has the capacity to use all the renewable energy resources existing in this region [12]. HOMER stands for Hybrid Optimization Model for Electric Renewables. Midwest Research Institute has the copyrights of this software [13]. It was developed by National Renewable Energy Laboratory (NREL) of United States. It is used to help the designing of various power plant configurations. Despite abundant availability of solar/wind energy, a PV or Wind Generator (WG) stand-alone system cannot satisfy the loads on a 24-hour basis [14]. Often, the variations of solar/wind energy generation do not match the time distribution of the load. Therefore, power generation systems dictate provision of battery storage facility to dampen the time distribution mismatch between the load and solar/wind energy generation and to facilitate for maintenance of the systems [15].

Using HOMER power optimization software for cost benefit analysis of hybrid-solar power generation relative to utility cost in Nigeria was studied by Ajao et al [16]. Techno-economic feasibility analysis of a solar PV grid-connected system with different tracking using HOMER software was investigated by Garni et al [17]. Aprillia studied the design On-Grid solar power system for 450 VA conventional housing using HOMER Software [18]. Desain sistem On-Grid energi Terbarukan Skala Rumah Tangga Menggunakan Perangkat Lunak HOMER was studied by aprillia [19].

In this paper, In this article, a 1 MW solar power plant in Sirjan city is studied using Homer software.

**Case Study**

Sirjan is a city and the capital of Sirjan County, Kerman Province, in the South of Iran. At 1730m, it is situated in a depression between the southern Zagros Mountains to the west and the Kuh-e Bidkhan massif to the east.

The curves of the solar radiation and wind speed for Sirjan city for each month in 2018 are presented in figures 1 and 2, respectively [3].

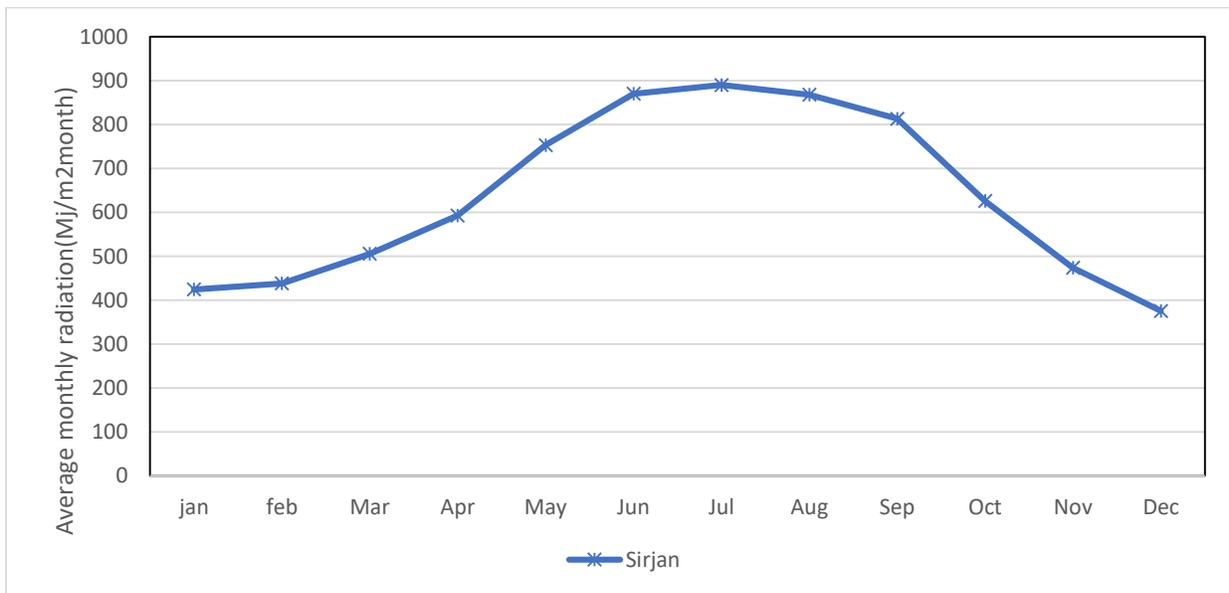

**Figure 1: Average monthly radiation (MJ/m$^2$) in Sirjan city [3]**

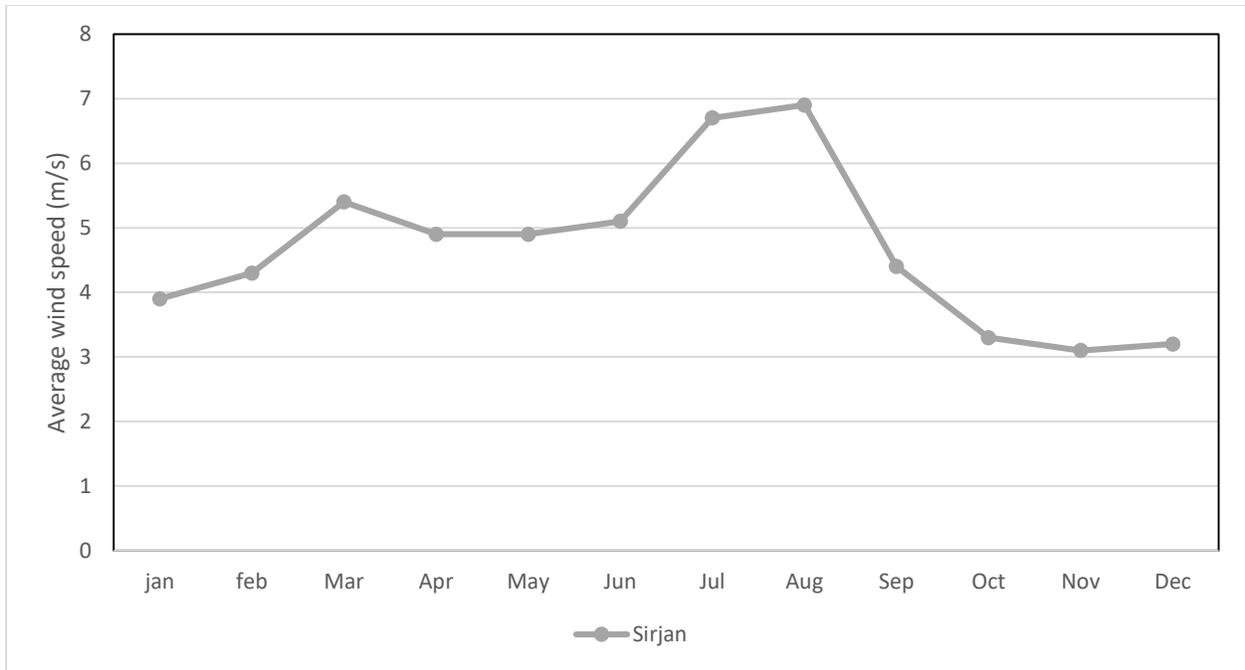

**Figure 2: Average wind speed (MJ/m$^2$) in Sirjan city [3].**

The following data have been extracted to build a 1 MW solar power plant in the Balord region of Sirjan. The design location of the solar power plant in Sirjan is shown in Figure 3.

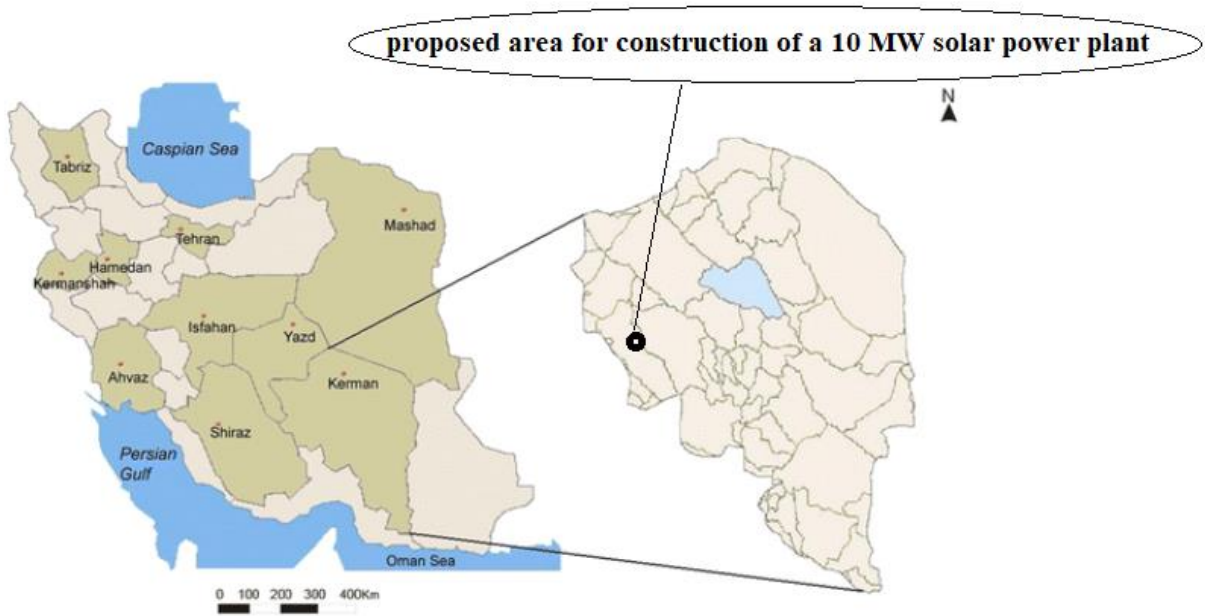

**Figure 3: Design location of the 1 MW solar power plant in Sirjan**

The design information of a 1 MW solar power plant in Sirjan city is given in Table 1.

**Table 1: Design information of a 10 MW solar power plant in Sirjan**

| Description | unit | amount |
|---|---|---|
| City | * | Sirjan |
| Longitude | North | 29 ° 6 'N |
| latitude | East | 58 ° 20' E |
| Power plant capacity | Megawatts | 1 |
| Area | square meters | 15,000 |
| Number of solar panels | number | 2, 500 |
| Dimensions of each panel | square meters | 2 |
| Cost of purchasing panels | US$ | 1,000,000 |
| The cost of building a power plant | US$ | 1,600,000 |
| Purchase price | US$ | 90,000 |
| Internal Rate of Return (IRR) | % | 18.05 |

**Results**

First, the geographical coordinates of Sirjan city are placed on Homer software. Also, discount rate, inflation rate, annual capacity shortage and project lifetime are 10 %, 5% , 5 % and 25 years, respectively. The peak of electricity consumption is seen in July. A schematic of the design can be seen in Figure 4.

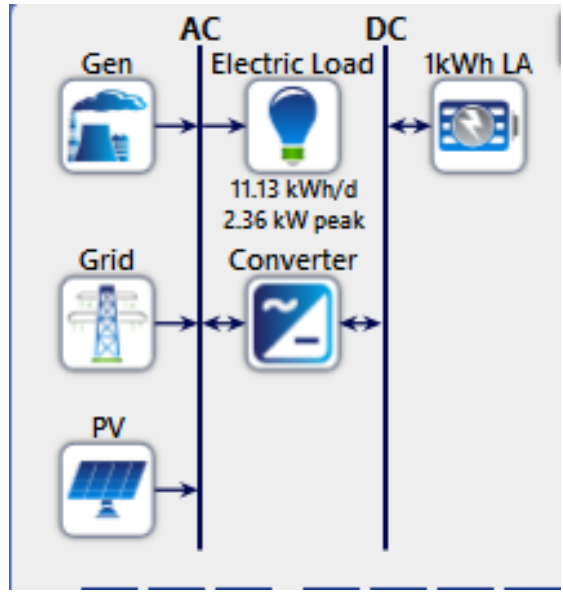

**Figure 4: Design location of the 1 MW solar power plant in Sirjan**

The AC load profile for load-shedding hours for one year is shown in Figure 5. It had a scaled annual average of 11.27 kWh/day and the peak load was 0.467 kW. Scaled data were used for calculations in HOMER. Only load values for load-shedding hours were used in the primary load inputs of HOMER; the rest of the fields in 24-h load profiles were set to zero. The average load shedding duration is 5h per day; however, this may vary accordingly with the load requirements of the consumers. The average monthly electricity production per day is shown in Figure 6. Also, the daily production per month is shown in Figure 7. This figure shows that the highest rate is in July.

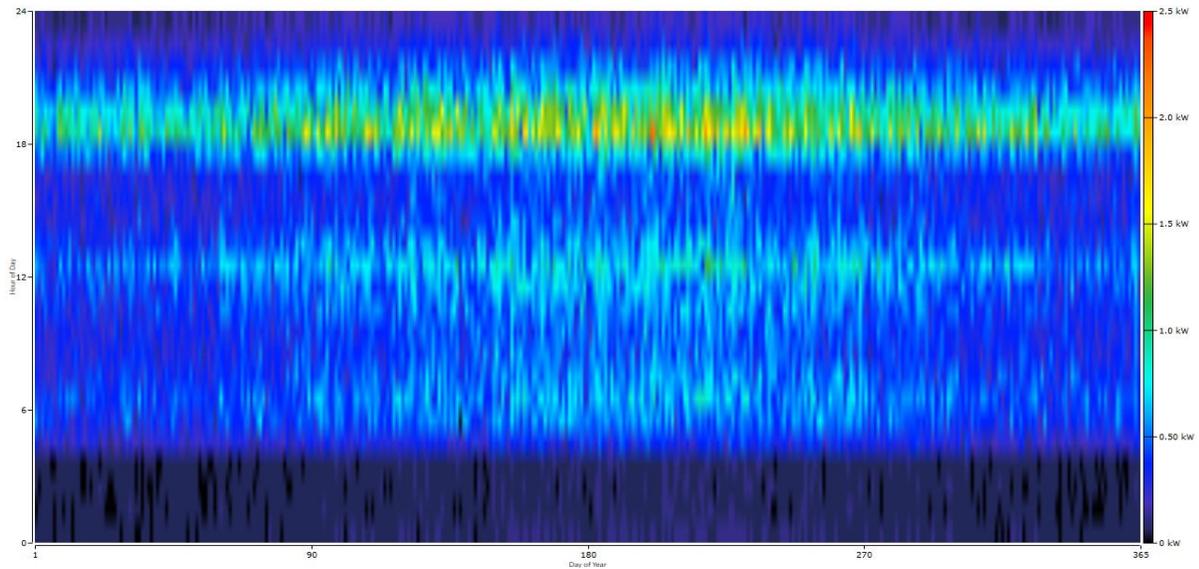

**Figure 5: The AC load profile for load-shedding hours for one year**

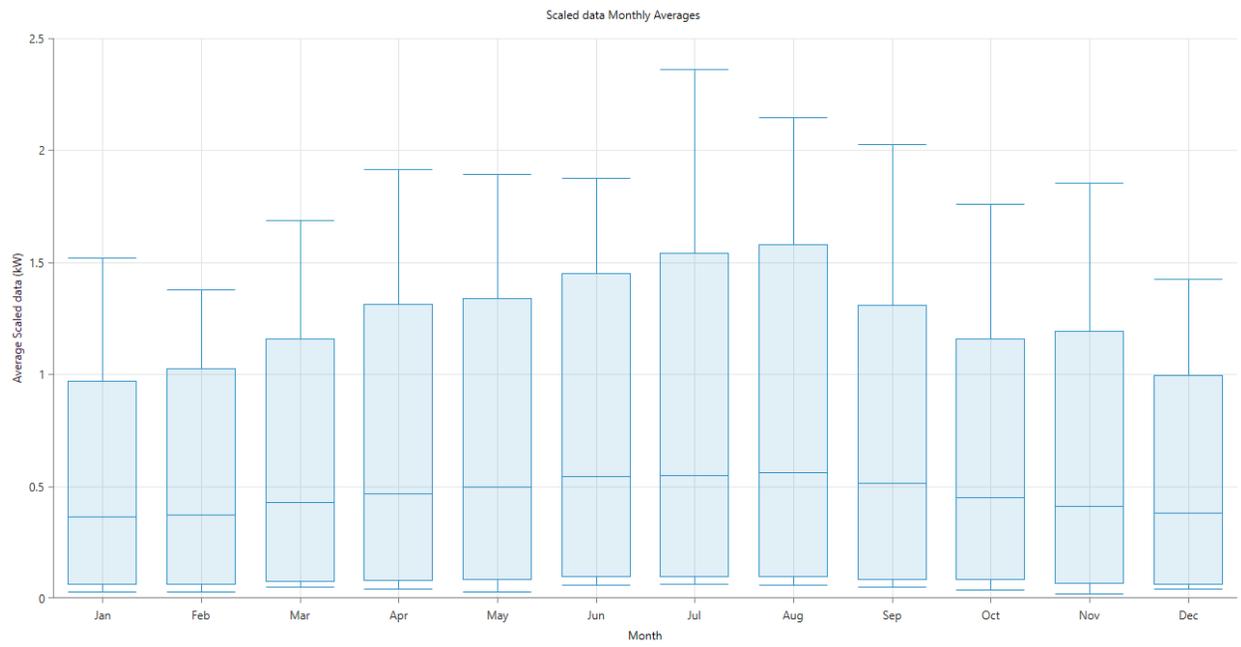

**Figure 6: The average monthly electricity production per month**

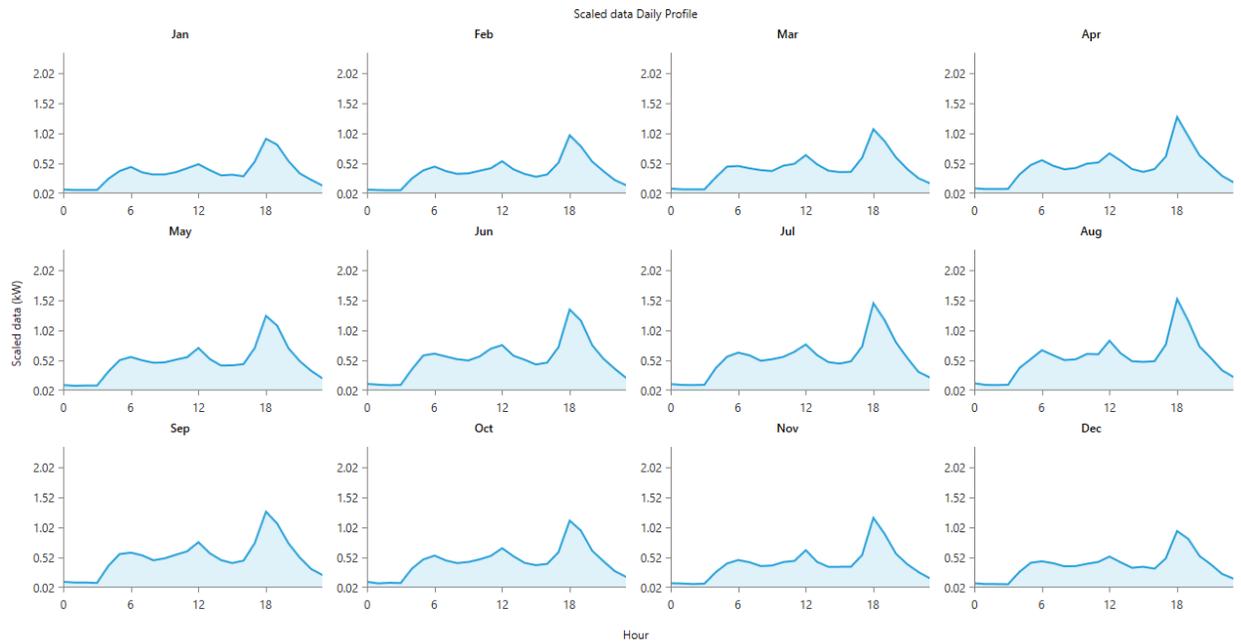

**Figure 7: The average monthly electricity production per month**

**Conclusion**

In this article, a 1 MW solar power plant was proposed to integrate with a diesel power plant of a local site in Sirjan, Iran. After investigating different case studies, it was concluded that Cost of purchasing panels and the cost of building a power plant were 1 M US$ and 1.6 M US$, respectively. For load-shedding hours, a PV system is suggested, which purchase costs 90000 US$. Discount rate, inflation rate, annual capacity shortage and project lifetime are 10 %, 5%, 5 % and 25 years, respectively. Scaled data were used for calculations in HOMER. It had a scaled annual average of 1127 kWh/day and the peak load was 0.467 kW. The maximum electricity energy is obtained in July.